\documentclass{article}

\usepackage{PRIMEarxiv}

\usepackage[utf8]{inputenc} 
\usepackage[T1]{fontenc}    
\usepackage{hyperref}       
\usepackage{url}            
\usepackage{booktabs}       
\usepackage{amsfonts}       
\usepackage{nicefrac}       
\usepackage{microtype}      
\usepackage{lipsum}
\usepackage{fancyhdr}       
\usepackage{graphicx}       
\graphicspath{{media/}}     

\newcommand{\rmnum}[1]{\romannumeral #1}
\newcommand{\tabincell}[2]{\begin{tabular}{@{}#1@{}}#2\end{tabular}}
\makeatother
\usepackage{newfloat}
\usepackage{listings}
\usepackage{amsmath}
\usepackage{multirow}
\usepackage[ruled,linesnumbered]{algorithm2e}
\usepackage{pifont}

\title{Multi-Modality Multi-Scale Cardiovascular Disease Subtypes Classification Using Raman Image and Medical History
\thanks{\textit{\underline{Citation}}: 
\textbf{Yu B, Chen H, Jia C, et al. Multi-modality multi-scale cardiovascular disease subtypes classification using Raman image and medical history[J]. Expert Systems with Applications, 2023: 119965.}} 
}

\author{
  Bo Yu \\
  School of Artificial Intelligence \\
  Jilin University \\
  Changchun, China\\
   \And
  Hechang Chen \\
  School of Artificial Intelligence \\
  Jilin University \\
  Changchun, China\\
  \texttt{chenhc@jlu.edu.cn} \\
  Corresponding Author \\
   \And
  Chengyou Jia \\
  Tongji University School of Medicine \\
  Tongji University \\
  Shanghai, China\\
   \And
  Hongren Zhou \\
  School of Artificial Intelligence \\
  Jilin University \\
  Changchun, China\\
   \And
  Lele Cong \\
  Department of Neurology \\
  China-Japan Union Hospital of Jilin University \\
  Changchun, China\\
  \texttt{congll18@mails.jlu.edu.cn} \\
  Corresponding Author \\
   \And
  Xiankai Li \\
  Tongji University School of Medicine \\
  Tongji University \\
  Shanghai, China\\
   \And
  Jianhui Zhuang \\
  Tongji University School of Medicine \\
  Tongji University \\
  Shanghai, China\\
   \And
  Xianling Cong \\
  Tissue Bank \\
  China-Japan Union Hospital of Jilin University \\
  Changchun, China\\
}

\begin{document}
\maketitle

\begin{abstract}
Raman spectroscopy (RS) has been widely used for disease diagnosis, e.g., cardiovascular disease (CVD), owing to its efficiency and component-specific testing capabilities.
A series of popular deep learning methods have recently been introduced to learn nuance features from RS for binary classifications and achieved outstanding performance than conventional machine learning methods.
However, these existing deep learning methods still confront some challenges in classifying subtypes of CVD. 
For example, the nuance between subtypes is quite hard to capture and represent by intelligent models due to the chillingly similar shape of RS sequences. 
Moreover, medical history information is an essential resource for distinguishing subtypes, but they are underutilized. 
In light of this, we propose a multi-modality multi-scale model called M3S, which is a novel deep learning method with two core modules to address these issues. 
First, we convert RS data to various resolution images by the Gramian angular field (GAF) to enlarge nuance, and a two-branch structure is leveraged to get embeddings for distinction in the multi-scale feature extraction module. 
Second, a probability matrix and a weight matrix are used to enhance the classification capacity by combining the RS and medical history data in the multi-modality data fusion module. 
We perform extensive evaluations of M3S and found its outstanding performance on our in-house dataset, with accuracy, precision, recall, specificity, and F1 score of 0.9330, 0.9379, 0.9291, 0.9752, and 0.9334, respectively. These results demonstrate that the M3S has high performance and robustness compared with popular methods in diagnosing CVD subtypes.
\end{abstract}


\section{Introduction}
Cardiovascular disease (CVD) is the leading cause of death globally, and coronary artery disease (CAD) is one of the main causes of mortality among CVD patients \cite{heart_disease1}.
Many studies indicate that patients with CAD tend to develop atrial fibrillation (AF), followed by acute myocardial infarction (AMI), which are the subtypes of CVD \cite{heart_disease2}.
Accurate classification of these subtypes should be performed as soon as possible, so appropriate therapy can be instituted to improve the survival rate.
In the past decade, methods like electrocardiograms, myocardial zymograms, cardiac computerized tomography scans, and coronary angiography are commonly used to diagnose them, which are time-consuming procedures \cite{heart_diagnosis}.
Nevertheless, diagnosis and treatment given within the first hour or `the golden hour' after a heart attack are crucial for patients, as this can save their life \cite{heart_book}.
Therefore, establishing an efficient method that can provide a timely and accurate diagnosis of these subtypes is an urgent task. 

Raman spectroscopy (RS) is regarded as a fast optical detection tool that can get unique features of fingerprints for recognition in minutes and requires only a tiny amount of blood, which has great significance for diagnosing CVD \cite{raman_heart}.
Meanwhile, as a modern detection technology, RS has the advantages of noninvasive tests, simple operation, and high sensitivity, which has achieved substantial success in the domains of agriculture \cite{agri1,agri2,agri3,agri4}, biology \cite{bio1,bio2}, chemistry \cite{che1,che2}, and medicine \cite{med1,med2,med3}. 
Based on these successful applications of RS for recognition above, the resulting ability to distinguish subtypes in CVD is reassuring.
Nevertheless, extracting valuable information from RS requires deep analyses by experienced experts or sophisticated statistical tools, limiting its applications adapted for the clinic. 
Fortunately, the fast development of artificial intelligence (AI) is becoming an efficient way to analyze RS. It can automatically extract and find minor spectral variations between species for classification instead of analysis based on their experience or statistical methods \cite{ai_raman1,ai_raman2}. 

The current solutions for a typical diagnosis by RS are typically based on conventional machine learning algorithms, e.g., discriminant analysis (DA), principal component analysis (PCA), support vector machine (SVM), etc \cite{ai_raman1}.
Specifically, Leong et al. use the partial least squares-discriminant analysis method (PLS-DA) to classify COVID-19 with RS data \cite{med1}. 
Most studies prefer to focus essentially on the variations between different diseases in RS data by using PCA and SVM \cite{machine1,machine2,machine4}. 
In addition, there also have some researchers use representative models such as decision tree (DT) \cite{machine7}, K-means clustering \cite{med2}, and random forest (RF) \cite{machine5} to analyze Raman data. 
Based on these studies, the latest works on RS use the PCA and LDA to make some new attempts for binary classification tasks on CVD \cite{machine8,machine9}.
Although RS has made some achievements in biomedical analyses with these methods, applying the technique to real-world applications is still challenging due to its shortcomings, such as sensitivity to noise and ease of overfitting.

Deep learning has recently gained popularity due to its ability to extract essential features automatically \cite{ai_raman2}.
For instance, representative models like artificial neural networks (ANN) \cite{deep4}, convolutional neural networks (CNN) \cite{deep1,deep2}, and long short-term memory networks (LSTM) \cite{deep6,deep7} are employed to extract high-dimensional features of RS data for subsequent diagnosis.
Compared with conventional machine learning methods, they usually perform better on binary classification tasks of RS due to their powerful representation capabilities.
And it has several other advantages, e.g., strong learning ability, good portability, robustness, etc., which make this high standard of application possible. 
However, methods that combine deep learning technologies and RS in publications do not provide new insight into CVD. So the goal of this study is to give a reference framework for this combination strategy to identify subtypes of CVD.

\begin{figure}[t]
\centering\includegraphics[width=0.5\textwidth]{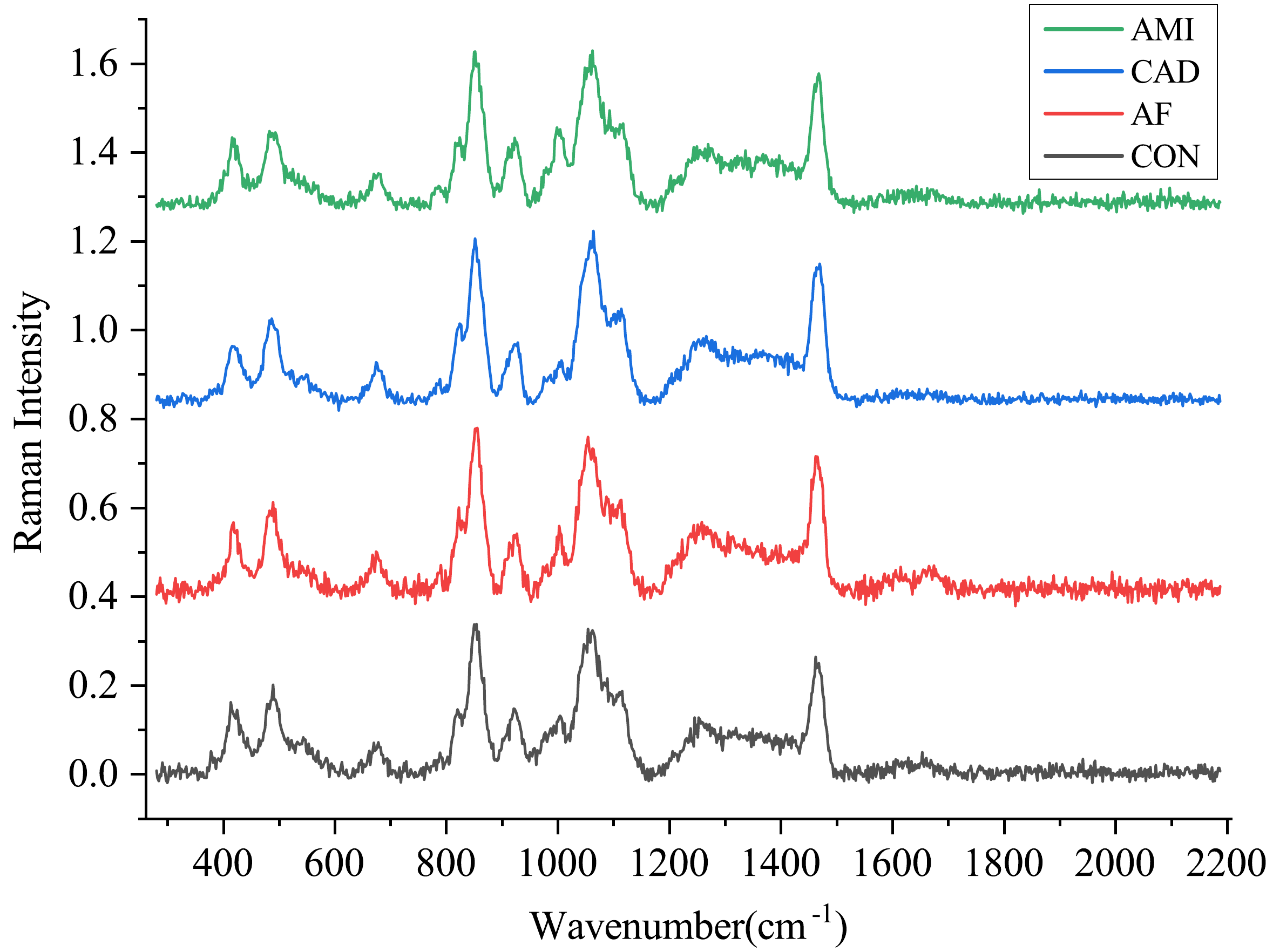}
\caption{The sampling schematic of Raman spectroscopy for four subtypes: acute myocardial infarction (AMI), coronary artery disease (CAD), atrial fibrillation (AF), control (CON).} 
\label{intro_raman}
\end{figure}

%
Despite the success that deep learning models have achieved in binary classification with RS, two significant challenges remain unsettled in CVD subtypes classification. 
\textbf{\emph{Challenge I: How to extract multi-scale features from RS?}}
Some investigators recently suggested using adaptive or diverse convolutional kernels to extract multi-scale features on one-dimensional (1D) RS sequences \cite{multiscale1,multiscale2}. 
However, as shown in Fig. \ref{intro_raman}, the shapes of RS sequences between subtypes and control groups are very similar. And it is laborious to discover their differences in 1D sequences by deep neural networks.
Therefore, capturing the multi-scale patterns from image perspectives, e.g., varies resolution and branch structure, will be potential ways to enlarge and represent these nuances to improve classification performance. 
\textbf{\emph{Challenge II: How to fuse multi-modality data beyond RS?}}
Recent works focus on combining optical coherence tomography (OCT) or fluorescence-based data with RS is an excellent way to improve diagnosis accuracy \cite{multimodal1,multimodal2}. 
However, collecting this additional data takes more time, which will waste time for rescue. 
On the contrary, the medical institution can quickly and conveniently obtain medical history information from patients during collecting patients' serum for Raman analysis. 
Unfortunately, this information which has a great reference value for diagnosing CVD subtypes in the clinic, is underutilized in existing models. 
So, incorporating Raman data and medical history information, i.e., multi-modality, is essential to a favorable outcome.

In view of this, we propose a multi-modality multi-scale (M3S) model, which incorporates the multi-scale feature extraction module and multi-modality data fusion module to address these challenges. 
Specifically, to enlarge nuances between different spectroscopies, we use the Gramian angular field algorithm (GAF) to convert RS data into images with different resolutions. 
And the dedicated multi-scale CNN is designed for feature extraction, which can articulate critical information in different resolution images using various kernels. 
The GAF allows us to map RS data into two-dimensional (2D) space to observe the details of images from different resolutions, and multi-scale CNN can effectively capture these features for subtypes classification in this module.
In the multi-modality data fusion module, we first establish a probability matrix based on the medical history information of patients, and it can represent the degree of dependence for diagnosis between medical histories and CVD subtypes. 
Then, a weight matrix with trainable parameters is generated to balance the importance between Raman data and medical history information in the decision-making stage. 
This strategy can fully use existing information, thereby helping the model improve classification ability through multi-modality data fusion. 
In summary, the contributions of this paper are as follows: 
\begin{itemize}
\item 
A novel model called M3S is proposed for CVD subtypes classification regarding Raman images and medical histories. It is the first model incorporating Raman data and medical history information, co-driving by the multi-scale feature extraction and multi-modality data fusion modules.

\item 
In the multi-scale feature extraction module, the RS sequence is converted into images of different resolutions by GAF, and a two-branch CNN feature extractor with different sizes of kernels is designed. This process can focus on subtle differences between subtypes and excellently describe them.

\item 
In the multi-modality data fusion module, the probability matrix is used to represent the degree of relevance between medical histories and RS sequences. And the weight matrix can give the distribution ratio between Raman data and medical history information during the decision-making process.

\item 
Extensive experimental results on in-house datasets compared with the state-of-the-art methods demonstrate the effectiveness of the M3S. Besides, more in-depth analyses are provided to explore the essential factors determining classification performance.
\end{itemize}

The remaining paper is organized as follows. 
Section 2 presents our solution for classifying CVD subtypes with the multi-modality multi-scale model. 
Experiments and detailed analysis are presented in Section 3, and Section 4 reviews previous work on RS application by existing AI technology. 
Finally, Section 5 covers the conclusion of the paper.

\section{Methodology}
In this section, the problem of CVD subtypes classification on 2D RS with medical history is defined in Section 2.1. We then propose a novel framework called M3S for this task in Section 2.2. Next, we elaborate on how to extract crucial information from multi-scale 2D RS by the convolutional neural network in Section 2.3. Finally, in Section 2.4,
we present the multi-modality combination algorithm, which incorporates the 2D RS predicted results and medical history indicated by two matrixes to predict the final subtypes.

\subsection{Problem Formulation}
We aim to construct a computer-aided diagnosis system based on two core phases, and they are defined as follows:

\emph{Phase I:} multi-scale feature extraction. Given a Raman sequence $S^{n}=\{s_{1}^{n}, s_{2}^{n}, ..., s_{1024}^{n}\}_{n=1}^{N}$ of $1024$ real-valued observations as the input, we need to rescale $S^{n}$ in the interval $[-1,1]$ and get a $\widetilde{S}^{n}$ that all values are fall in this domain. The current sample serial number is $n$, and the total sample size of the training set or testing set is $N$, $n\in N$. The sequence size of $\widetilde{S}^{n}$ is also downsample to $i$ groups by the piecewise aggregation approximation algorithm (PAA) at the same time. After that, the new sequence is converted to a polar coordinate, and ${\phi}^{n}_{i}$ represents the angle value of each point. The GAF algorithm will calculate the cosine relationship between different ${\phi}^{n}_{i}$ and get a 2D image $A_{i}^{n}$, which is we need with scale $i$ of the $n^{th}$ sample. Finally, we design two CNN structures called $G_{f1}$ and $G_{f2}$ with different kernel sizes to extract the features $e_{1}^{n}$ and $e_{2}^{n}$ from multi-scale $A_{i}^{n}$. And there has a fully connection (FC) layer $G_{l}$ can give the preliminary prediction $e_{R}^{n}$ by $e_{3}^{n}$, which is formed by concatenate $e_{1}^{n}$ and $e_{2}^{n}$.

\emph{Phase II:} multi-modality data fusion. First, we collect five categories of medical history information corresponding to $N$ samples and then build a probability matrix $M_{H}$. Second, the last fusion features $e_{F}^{n}$ are organized by concatenating the classification result $e_{R}^{n}$ in the prior stage and the $M_{H}$. Third, we design a weight matrix consisting of trainable parameters $M_{W}$, and it is multiplied by the $e_{F}^{n}$ to get the prediction matrix $P_{m}^{n}$. Finally, we sum $P_{m}^{n}$ by classes and use the softmax function to predict subtypes $P^{n}$ with the highest probability. The parameters of the neural network are trained by calculating the loss value between $P^{n}$ and $Y^{n}$, where $Y^{n}$ represents the subtypes label of $n^{th}$ sample.

\begin{table}[t]
\centering
\caption{Key notations used in this paper.}
\label{symbols}
\begin{tabular}{c|p{8cm}}
\toprule
Symbol & Meaning \\
\midrule
$S^{n}$ & Raman wavenumber sequences \\
$\widetilde{S}^{n}$ & sequences that have been rescaled \\
$N$ & sample size of the training set or test set \\
${\phi}^{n}_{i}$ & spectral value in the polar coordinates \\
$A_{i}^{n}$ & Raman image \\
$G_{f}$ & feature extractor by CNN \\
$G_{l}$ & predictor by FC\\
$e_{R}^{n}$ & classification result with Raman image \\
$e_{F}^{n}$ & last fusion features \\
$M_{H}$ & training or test probability matrix \\
$M_{W}$ & weight matrix \\
$P_{m}^{n}$ & prediction matrix \\
$P^{n}$ & subtypes prediction of sample \\
$Y^{n}$ & subtypes label of sample \\
\bottomrule
\end{tabular}
\end{table}

In general, given a 1D spectroscopy $S^{n}$, we transform them to an image $A_{i}^{n}$ by GAF, and the CNNs are leveraged to extract crucial features from them. The model can give final results $P^{n}$ according to the features that confusing the classification result $e_{R}^{n}$ and medical history $M_{H}$ by a weight matrix $M_{W}$. An overview of the notations used in this paper is provided in Table \ref{symbols}.

\begin{figure*}[t]
\centering\includegraphics[width=0.95\textwidth]{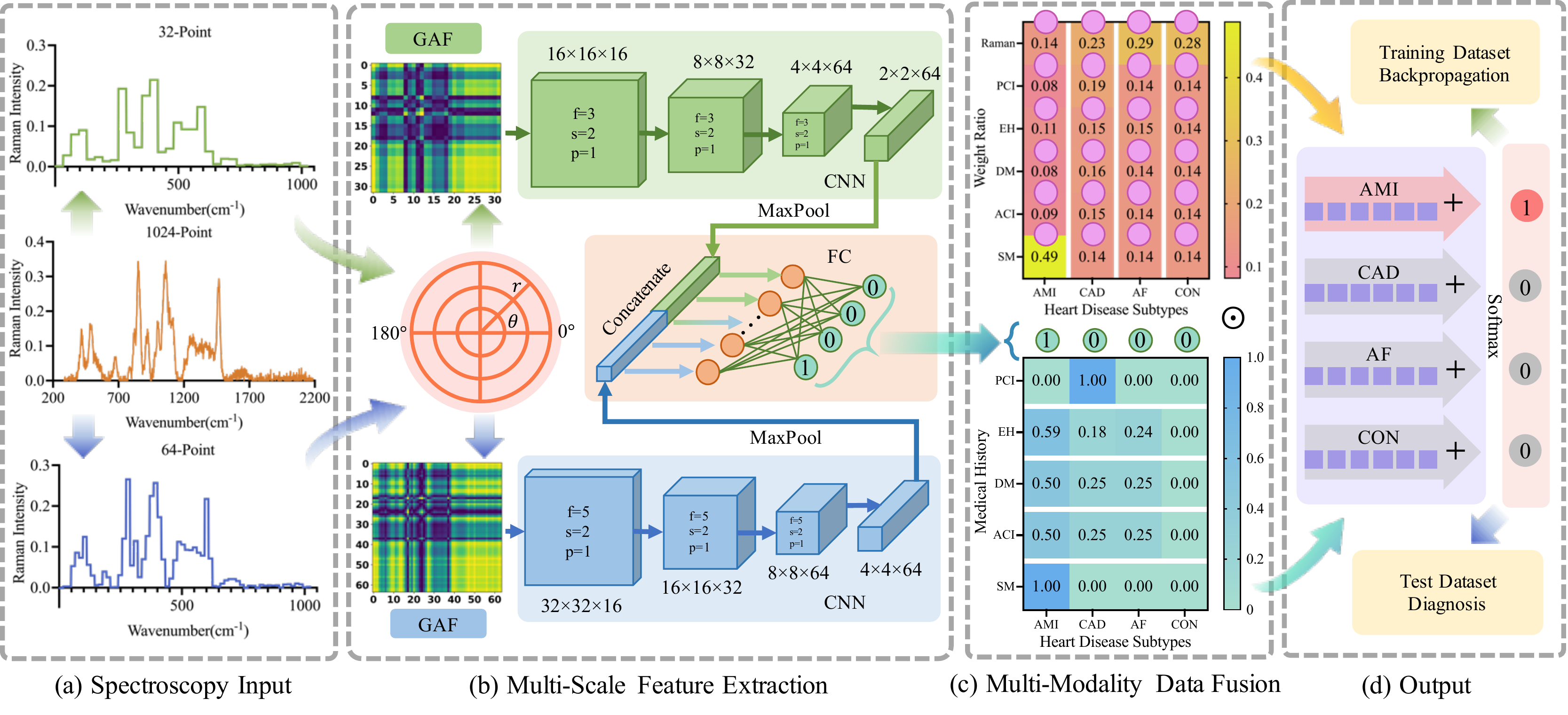}
\caption{Overview of M3S architecture. In (a), the data in the middle is raw, and different sampling rates obtain the upper and lower data. In (b), $f$ represents the kernel size, $s$ represents the stride, and $p$ represents the padding. The last convolutional layer parameters are: $f=2$, $s=2$, $p=0$. In (c), the corresponding positions of the probability matrix and the weight matrix are multiplied. In (d), subtypes classification results are obtained by weighted summation of the prediction matrix.} 
\label{architecture}
\end{figure*}

\subsection{The Framework of M3S Model}
Fig. \ref{architecture} shows the four parts that mainly make up the M3S: (a) spectroscopy input, (b) multi-scale feature extraction, (c) multi-modality data fusion, and (d) output.

\emph{Part \rmnum{1}:} We use 1D RS data from two different sampling frequencies as input (Fig. \ref{architecture} (a)). In this paper, the groups $i$ are 32 and 64, respectively.

\emph{Part \rmnum{2}:} We map the resampled RS data into a polar coordinate and then use the GAF algorithm to generate the corresponding 2D Raman images for features extracted with CNN. An FC layer is used at the end of the model to generate preliminary predictions at this stage (Fig. \ref{architecture} (b)). 

\emph{Part \rmnum{3}:} The model counts the medical history of the training or testing samples and generates a medical history probability matrix combined with preliminary predictions, and a weight matrix consisting of neuron parameters is trained. It will be multiplied by the probability matrix to get the prediction matrix (Fig. \ref{architecture} (c)). 

\emph{Part \rmnum{4}:} The model sums the prediction matrix along the distribution direction of the subtypes and uses the softmax function to give the final prediction. It will enforce the diagnosis process during this phase (Fig. \ref{architecture} (d)). 

Next, we will describe in detail for \emph{Part \rmnum{2}} and \emph{Part \rmnum{3}}, which are the two essential components of M3S.

\subsection{Multi-Scale Feature Extraction}
Feature extraction aims to find the most condensed and informative set of embeddings to enhance the performance of the model. Furthermore, it is employed to extract features from the original signal, enabling the algorithm to gain some discrimination abilities. Therefore, feature extraction is a critical step in the modeling process, and we used two efficient techniques to build it. First, since the 1D Raman spectroscopies of CVD subtypes are very similar, we leverage GAF to convert them into 2D signals, e.g., Raman image, which can provide more information to assist classification. Second, we design a multi-scale CNN feature extractor for mining the valuable information for classification in Raman images. It is composed of two branches with different convolution kernel sizes, and each branch can extract different levels of potential features for classifying subtypes. 
Algorithm \ref{algorithm1} describes the feature extraction process of M3S.

\begin{algorithm}[tb]
    \KwIn{$S^{n}$}
    \KwOut{$e_{R}^{n}$}
    Initialization : $G_{f1}$, $G_{f2}$, $G_{l}$\;
    \% Training for E epoch\\
    \For{e = 1 to E}
        {\For{n = 1 to N}
            {
            \% Raman Gramian Angular Field \\
            $\widetilde{S}^{n}=Rescale(S^{n})$\;
            $A_{i}^{n}=GAF(\widetilde{S}^{n})$\;
            \% Multi-Scale Extractor \\
            $e_{1}^{n}=G_{f1}(A_{i=32}^{n},\theta_{f1})$ \; 
            $e_{2}^{n}=G_{f2}(A_{i=64}^{n},\theta_{f2})$ \;
            $e_{3}^{n}=Concat(Maxpool(e_{1}^{n}),Maxpool(e_{2}^{n}))$ \;
            $e_{R}^{n}=G_{l}(e_{3}^{n},\theta_{l})$ \;
            }
        }
  \caption{Multi-Scale Feature Extraction}
  \label{algorithm1}
\end{algorithm}

\begin{figure}[t]
\centering\includegraphics[width=0.4\textwidth]{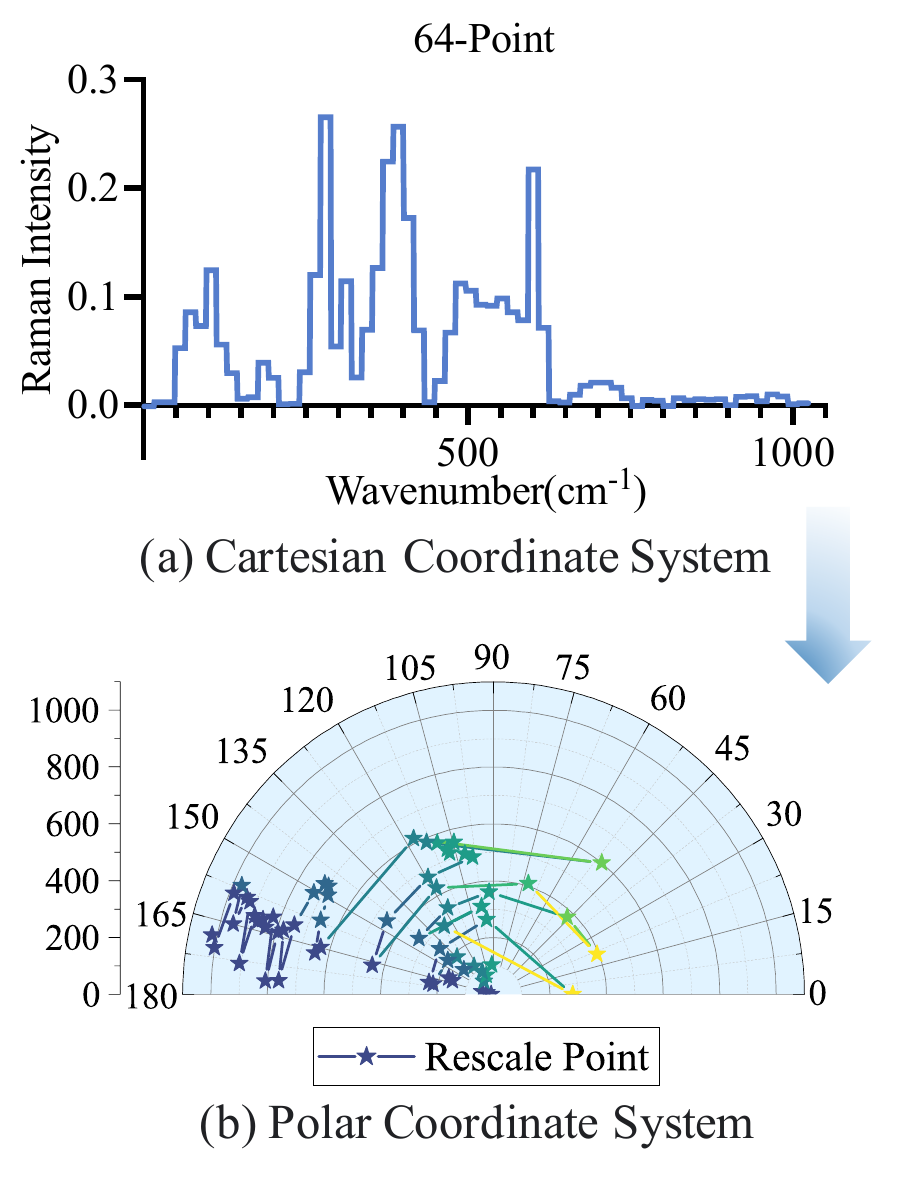}
\caption{The distribution of RS in polar coordinates after resampling. It shows the mapping process of the AMI data resampled to 64 points from the Cartesian coordinate system (a) to the polar coordinate system (b).} 
\label{polar}
\end{figure}

\textbf{Raman Gramian Angular Field.} 
Inspired by successes using deep learning in computer vision and natural language processing, Wang et al. propose a framework to encode time sequence as a type of image called Gramian Angular Field (GAF) \cite{gaf}. By using the polar coordinate system, GAF images are represented as a Gramian matrix where each element is the trigonometric sum between different sample points. For the $n^{th}$ sample in our model, the Raman wavenumber sequences can be represented as $S^{n}=\{s_{1}^{n}, s_{2}^{n}, ..., s_{1024}^{n}\}_{n=1}^{N}$ with $1024$ real-valued points. The spectral needs to be rescaled in the range of $[-1,1]$, and it is obtained by
\begin{equation}
\widetilde{S}^{n}_{k}=\frac{({S}^{n}_{k}-MAX(S^{n}))+({S}^{n}_{k}-MIN(S^{n}))}{MAX(S^{n})-MIN(S^{n})},
\end{equation}
where $k$ represents the order of $S^{n}$. As result, we get a normalized sequences $\widetilde{S}^{n}$ that all values are distributed between $-1$ and $1$. This step will use the PAA algorithm to divide a $1024$ sequence into $i$ groups in the meantime, and the number of groups $i$ is the length of the new sequence. Next, it can represent the rescaled RS $\widetilde{S}^{n}$ in polar coordinates by encoding the value as the angular cosine ${\phi}^{n}_{i}$ and timestamp as the radius $r^{n}$ with
\begin{equation}
\begin{cases}
     {\phi}^{n}_{i}=arccos(\widetilde{S}^{n}_{i}), & -1\leq \widetilde{S}^{n}_{i}\leq 1 \\  
     r^{n}_{i}=\frac{t_{i}}{B}, & t_{i}\in B,
\end{cases}
\end{equation}

where $t_{i}$ represents the time stamp of group $i$, $B$ is a constant factor to regularize the span of the polar coordinate system. Fig. \ref{polar} shows the process of data mapping. Next, we will generate Raman images by combining these data with GAF, which has a default setting called "Summation" as 
\begin{equation}
A_{i}^{n}=
\begin{bmatrix} 
cos({\phi}^{n}_{1}+{\phi}^{n}_{1}) & \dots & cos({\phi}^{n}_{1}+{\phi}^{n}_{i}) \\ 
\dots & \dots & \dots \\
cos({\phi}^{n}_{i}+{\phi}^{n}_{1}) & \dots & cos({\phi}^{n}_{i}+{\phi}^{n}_{i})
\end{bmatrix},
\end{equation}
where the GAF can identify the temporal correlation with different groups by considering the trigonometric sum between each point and deeply exploiting the angular perspective ${\phi}^{n}_{i}$. After that, we can get a Raman image $A_{i}^{n}$ that we need with the scale $i$ of the $n^{th}$ sample.

\textbf{Multi-Scale Extractor.} 
The GAF can generate Raman images of corresponding resolutions according to the size of sampling points, and there has a variety of information in different resolution images. In computer vision, high-resolution images contain more detailed information, whereas low-resolution images contain more abstract knowledge. Furthermore, in a Raman image, high-resolution retains more local cosine relationships between points, while low-resolution expresses more global semantic connections between different groups. Meanwhile, with the development of deep learning, CNN has become the most popular application to analyze the image in computer vision \cite{ai_medical_image1,ai_medical_image2}. Therefore, this paper chooses Raman images with two resolutions of 32 and 64 as input and designs a multi-scale CNN feature extraction network in a targeted manner.

\begin{figure}[t]
\centering\includegraphics[width=0.5\textwidth]{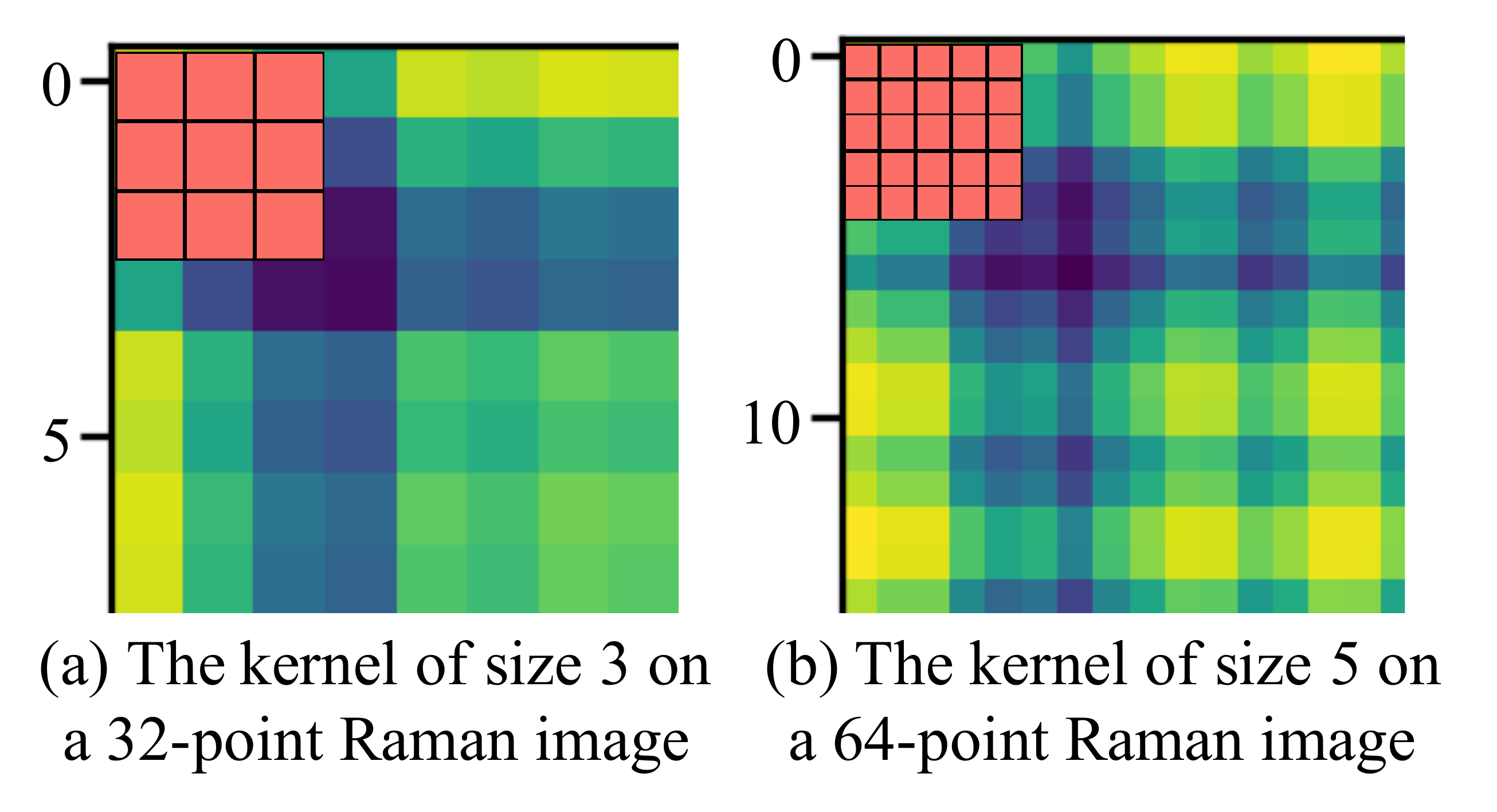}
\caption{Variation of convolution kernel size on Raman images of different resolutions.} 
\label{kernel}
\end{figure}

In order to keep the receptive field of the convolution operation roughly balanced on these two Raman images, we use a convolution network $G_{f1}$ to keep the kernel size at 3 on the 32-point image and another convolution network $G_{f2}$ to keep the kernel size at 5 on the 64-point image. $G_{f1}$ and $G_{f2}$ can be performed as
\begin{equation}
\begin{cases}
     e_{1}^{n}=G_{f1}(A_{i=32}^{n},\theta_{f1}) \\  
     e_{2}^{n}=G_{f2}(A_{i=64}^{n},\theta_{f2}),
\end{cases}
\end{equation}
where the $e_{1}^{n}$ and $e_{2}^{n}$ represent the embeddings in different dimensions, and we use 32 and 64 sizes in this paper, respectively. In Fig. \ref{kernel} (a), the abstract information is more prominent, and the convolution operation can extract the semantic information. In Fig. \ref{kernel} (b), the detailed information is more intact, which can help the extractor find the subtle characteristic of different diseases. The structure with two sizes of convolution kernels can roughly maintain the balance of the receptive fields on images with different resolution. Finally, the features of these two scales are concatenated into one embedding $e_{3}^{n}$, and a fully connected layer $G_{l}$ is used to give $e_{R}^{n}$ by
\begin{equation}
     e_{R}^{n}=G_{l}(e_{3}^{n},\theta_{l}),
\end{equation}
where the $\theta_{f1}$, $\theta_{f2}$, and $\theta_{l}$ are the parameters of neural networks, which are need to be optimized. The $e_{R}^{n}$ is the classification result from the multi-scale module without any assistance from medical history. That is why we call it the preliminary prediction.

\begin{algorithm}[t]
    \KwIn{$e_{R}^{n}$}
    \KwOut{$P^{n}$}
    Initialization : $M_{W}$\;
    \% Training for E epoch\\
    \For{e = 1 to E}
        {\For{n = 1 to N}
            {
            \% Probability Matrix \\
            $M_{H}\leftarrow Medical\  history$ \;
            $e_{F}^{n}=Concat(e_{R}^{n},M_{H})$ \;
            \% Weight Matrix \\
            $P^{n}_{M}=e_{F}^{n}\cdot M_{W}$ \; 
            $P^{n}=Softmax(SUM(P^{n}_{M}, axis=History), axis=Subtypes)$ \;
            }
        }
  \caption{Multi-Modality Data Fusion}
  \label{algorithm2}
\end{algorithm}

\subsection{Multi-Modality Data Fusion}
In information representation, the word "modality" can be defined as a medium that disseminates information. 
Accordingly, "multi-modality" can be defined as a method that generally uses more than one medium in information representation and propagation. 
In other words, multi-modality allows an integrated usage of various forms to interact closely in one system, which is widely used for data fusion in medicine, e.g., PET-CT and PET-MR with single nucleotide polymorphism (SNP) data \cite{multimodal4,multimodal3}. 
It can provide multiple information on the patient between images, text, etc., greatly assisting doctors in analyzing the disease from a broad perspective. Besides, the medical history is an essential basis for diagnosing and treating. 
Thus we hope to combine the medical history and RS images in a multi-modality manner to improve subtypes classification performance. 
To effectively fuse these two kinds of data, we create the probability and weight matrix with different functions at this stage. 
First, we obtain a probability matrix by performing statistical analysis on the medical history of the samples, which is used to express the determining factor of each medical history on the subtypes classification. 
Second, a weight matrix is designed by us that can assign the mixing ratio between the preliminary predictions from Raman images and medical history from the probability matrix. Algorithm \ref{algorithm2} describes the data fusion process of M3S.

\begin{figure}[t]
\centering\includegraphics[width=0.5\textwidth]{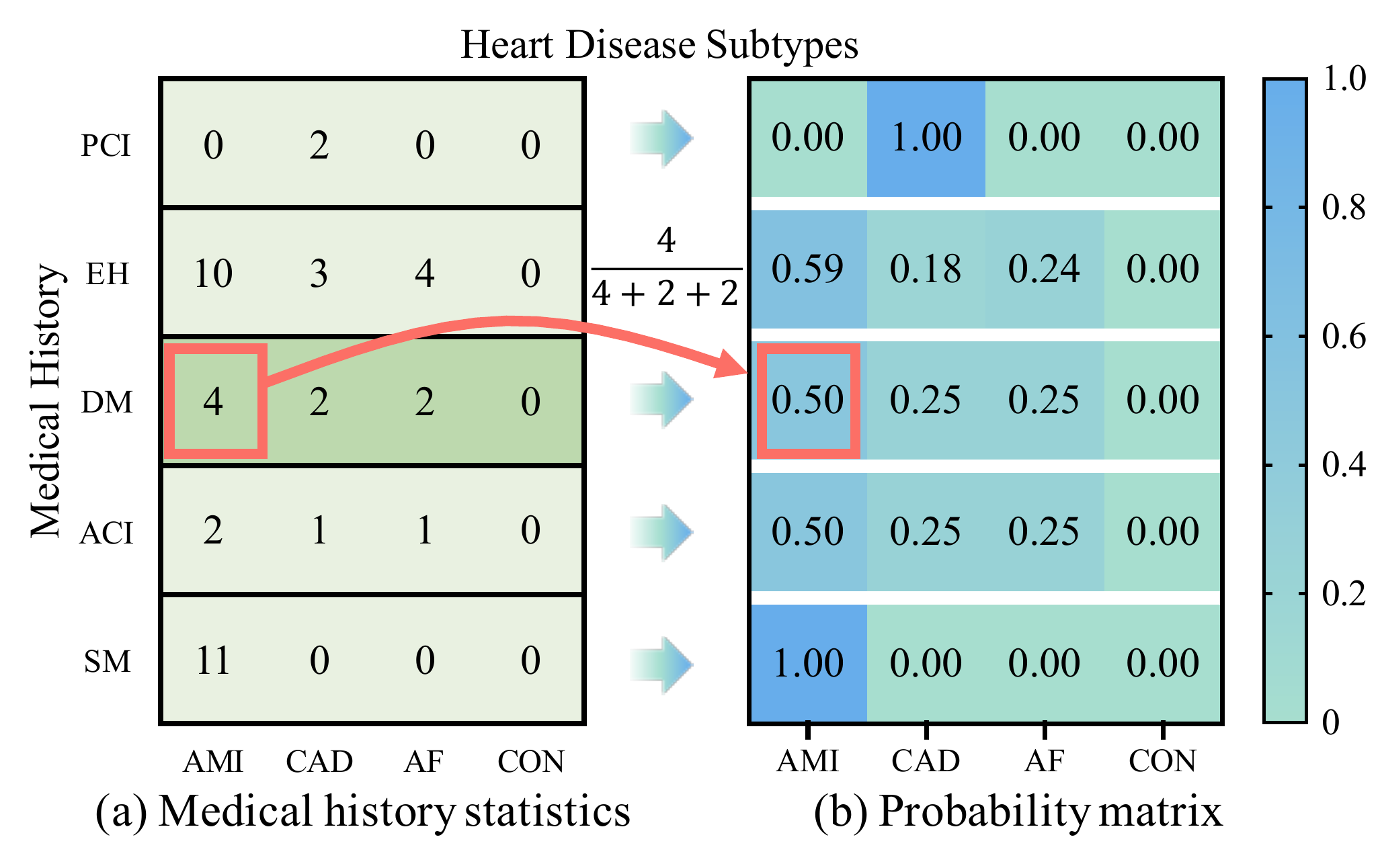}
\caption{The calculation process of the probability matrix.} 
\label{probability_matrix_image}
\end{figure}

\textbf{Probability Matrix.} 
After reviewing the relevant literature and consulting experts \cite{heart_disease1,heart_disease2,heart_book,heart_diagnosis}, we select five medical histories as reference indicators to aid the diagnosis of CVD subtypes: percutaneous coronary interventon (PCI), essential hypertension (EH), diabetes mellitus (DM), acute cerebral infarct (ACI), and smoking (SM). We collect medical history information from patients and enumerate this information in detail when participants sign informed consent agreements. As shown in Fig. \ref{probability_matrix_image} (a), we count the number of samples for each CVD subtype in the range of one medical history. Then, we divide the number of one subtype by the total number of this medical history, and we can get the probability belongs to the corresponding disease like Fig. \ref{probability_matrix_image} (b). All the calculated probabilities compone the probability matrix $M_{H}$ that we need, and it can indicate the degree of influence of each medical history on subtypes diagnosis. It should be noted that the probabilities are from all datasets for clear expression in Fig. \ref{probability_matrix_image}, but we can only use the medical history information in the training dataset.

\begin{figure}[t]
	\centering\includegraphics[width=0.5\textwidth]{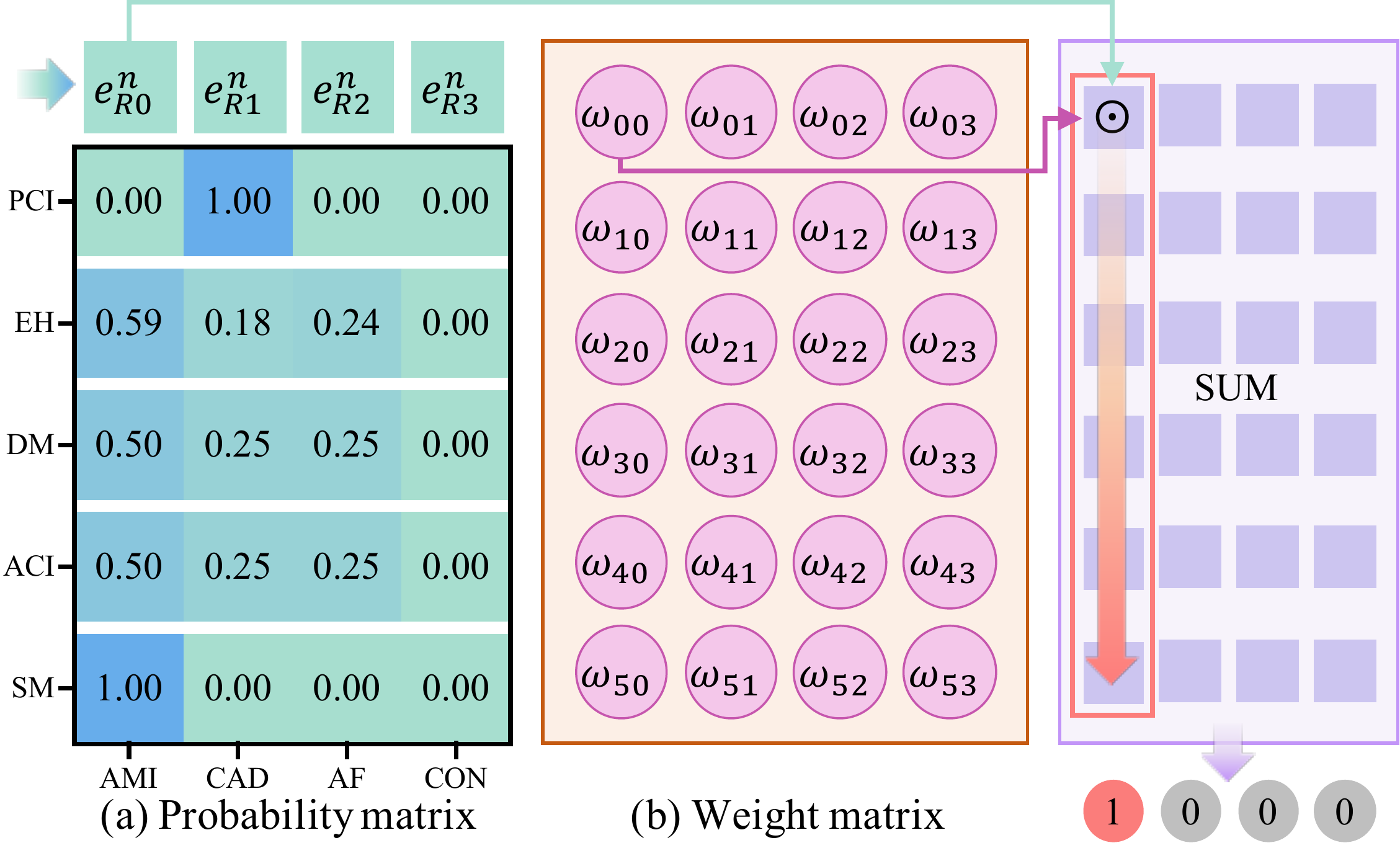}
	\caption{The calculation process of the weight matrix.} 
	\label{weight_matrix_image}
\end{figure}

\textbf{Weight Matrix.} 
Given the preliminary predictions from Raman images and a probability matrix of the medical history, how should we combine these two modalities of data for subtypes classification? Here, we design a weight matrix $M_{W}$, where each parameter $\omega$ represents the mixing ratio between RS data and medical histories for classifying subtypes. These parameters will be initialized at the beginning and trained along with the feature extractor to get the final mixing ratio. In view of this, M3S is an end-to-end classification system with multi-scale and multi-modality modules. As shown in Fig. \ref{weight_matrix_image}, we first concatenate the preliminary prediction results $e_{R}^{n}$ and the probability matrix $M_{H}$ to get a feature group $e_{F}^{n}$, and initialize a weight matrix $M_{W}$ with the same size of it. Next, multiplying feature group $e_{F}^{n}$ by weight matrix $M_{W}$, and the prediction matrix $P^{n}_{M}$ is obtained by
\begin{equation}
    P^{n}_{M}=e_{F}^{n}\cdot M_{W}.
\end{equation}
Finally, we sum data in $P^{n}_{M}$ by a medical history dimension and then use the softmax activation function by the subtypes dimension to get the final prediction result $P^{n}$.

\section{Experiments}
In this section, we first introduce some details about the dataset, evaluation methods, and hyperparameters used in this paper. 
Then we conduct four groups of experiments to answer the following questions: 
\textbf{Q1}: How does our proposed method perform compared with the state-of-the-art methods? 
\textbf{Q2}: How are different resolutions of RS images transformed and represented with the multi-scale feature extraction module?  
\textbf{Q3}: Is it necessary to automatically combine medical history information with RS data in the multi-modality data fusion module? 
\textbf{Q4}: What are the advantages and significance of M3S for classifying subtypes of CVD? 

\subsection{Implementation Details}
\subsubsection{Dataset}
Ultrafiltration serum samples are collected from Shanghai Tenth People's Hospital in 2021, including 31 CVD samples (18 AMI, 8 CAD, and 5 AF). And the control group (CON) is composed of 16 healthy volunteers without any major diseases. 
In order to further increase the amount of data available, the experiment collects dozens of spectroscopies from different points on each serum sample, so there have 1489 RS sequences (227 CAD, 546 AMI, 201 AF, and 515 CON). 
We also obtain five medical history information (PCI, EH, DM, ACI, SM) from the medical files of these patients. 
Before the collection of serum samples and medical files, informed consent and ethical approval are acquired from all patients. 
All experiments are performed in accordance with relevant laws and regulations. 
The protocol of the study is approved by the Jilin University and Shanghai Tenth People's Hospital, respectively.

The 10 kD serum samples from subjects are collected and measured on a WITec alpha300R Raman microscope (Oxford Instruments Group, England).
The laser power on the sample is 10 - 14 mW, and a grating with 1200 g/mm is used to describe the Raman sequence for recording. 
A 532 nm laser for Raman measurement is used, the single-spectrum integration time is 16 - 20 s, and the excitation light is focused using a 100X objective (NA = 0.9, Zeiss).
The spectroscopy range is 280 – 2187 $cm^{-1}$, and there are 1024 points.
The spectral process procedure of samples includes subtraction of background spectra, baseline calibration, and normalization.
The data are randomly divided into two parts, of which 75\% is classified as the training set, and 25\% is classified as the testing set. 

\subsubsection{Baselines and Evaluation}
To comprehensively compare the performance between our method and the state-of-the-art algorithms, we choose thirteen outstanding algorithms on RS analysis from two groups of baselines to verify. 
The public codes of these algorithms are used from the scikit-learn library or GitHub, and we reproduced the code for the unpublished works. 

\begin{itemize}
\item\emph{Conventional machine learning methods}:
\end{itemize}

(1) PCA-linear discriminant analysis (LDA) \cite{machine9}, 

(2) PCA-quadratic discriminant analysis (QDA) \cite{machine6}, 

(3) PCA-RF \cite{machine5}, 

(4) PCA-DT \cite{machine7}, 

(5) PCA-k nearest neighbors (KNN) \cite{machine10}, 

(6) PCA-Adaboost \cite{machine7}, 

(7) PCA-SVM \cite{machine2}

\begin{itemize}
\item\emph{Deep learning methods}:
\end{itemize}

(1) ANN \cite{deep4},

(2) CNN \cite{deep2},

(3) LSTM \cite{deep6},

(4) CNN-recurrence plot (RP) \cite{raman_image2},

(5) CNN-Gramian angular field (GAF) \cite{raman_image2},

(6) CNN-Markov transition field (MTF) \cite{raman_image2}

We used five evaluation metrics for test: accuracy (ACC), precision (P), recall (R, the same as Sensitivity), specificity (S), and F1 score (F1). Besides, we use floating point operations (FLOPs) and parameters (Params) metrics to compare the complexity of the deep learning models.

\begin{table}[t]
\centering
\caption{Performance comparison between our model and baselines on the CVD subtypes identification task.}
\footnotesize
\resizebox{\textwidth}{70mm}{
\begin{tabular}{cccccc|c|cccccc}
\toprule
\multicolumn{6}{c|}{Conventional machine learning methods} & & \multicolumn{6}{c}{Deep learning methods}\\
\midrule
Method & ACC & P & R & S & F1 & Items & ACC & P & R & S & F1 & Method\\
\midrule
\midrule
\multirow{6}{*}{\tabincell{c}{PCA-LDA\\\cite{machine9}}}& 0.7373 & 0.7429 & 0.7006 & 0.9022 & 0.7138 & 1 & 0.8231 & 0.8202 & 0.8049 & 0.9367 & 0.8115 & \multirow{6}{*}{\tabincell{c}{ANN\\\cite{deep4}}}\\
		 & 0.7024 & 0.7041 & 0.6683 & 0.8903 & 0.6817 & 2 & 0.8231 & 0.8290 & 0.8135 & 0.9350 & 0.8205 &\\
		 & 0.7131 & 0.7175 & 0.6808 & 0.8931 & 0.6885 & 3 & 0.7560 & 0.7596 & 0.7432 & 0.9136 & 0.7505 &\\
		 & 0.6997 & 0.7130 & 0.6642 & 0.8873 & 0.6789 & 4 & 0.8150 & 0.8075 & 0.8006 & 0.9346 & 0.8034 &\\
		 & 0.6863 & 0.6913 & 0.6487	& 0.8834 & 0.6619 & 5 & 0.8445 & 0.8525 & 0.8192 & 0.9419 & 0.833 &\\
		 \specialrule{0em}{0.5pt}{0.5pt}
		 \cline{2-12}
		 \specialrule{0em}{1pt}{1pt}
		 & 0.7078 & 0.7138 & 0.6725 & 0.8913 & 0.6850 & Average & 0.8123 & 0.8138 & 0.7963 & 0.9323 & 0.8038 &\\
\midrule
\multirow{6}{*}{\tabincell{c}{PCA-QDA\\\cite{machine6}}}& 0.7802 & 0.7887 & 0.7972 & 0.9211 & 0.7879 & 1 & 0.8150 & 0.7883 & 0.7863 & 0.9316 & 0.7871 & \multirow{6}{*}{\tabincell{c}{CNN\\\cite{deep2}}}\\
		 & 0.7587 & 0.7761 & 0.7947 & 0.9160 & 0.7716 & 2 & 0.8016 & 0.8056 & 0.7796 & 0.9278 & 0.7872 &\\
		 & 0.7587 & 0.7680 & 0.7759 & 0.9136 & 0.7665 & 3 & 0.8097 & 0.8115	& 0.7845 & 0.9302 & 0.7944 &\\
		 & 0.7748 & 0.7669 & 0.7843 & 0.9213 & 0.7709 & 4 & 0.8123 & 0.8207 & 0.8210 & 0.9328 & 0.8203 &\\
		 & 0.7828 & 0.7864 & 0.8104 & 0.9228 & 0.7929 & 5 & 0.7989 & 0.8070 & 0.7979 & 0.9290 & 0.8005 &\\
		 \specialrule{0em}{0.5pt}{0.5pt}
		 \cline{2-12}
		 \specialrule{0em}{1pt}{1pt}
		 & 0.7710 & 0.7772 & 0.7925 & 0.9189 & 0.7780 & Average & 0.8075 & 0.8066 & 0.7938 & 0.9303 & 0.7979 &\\
\midrule
\multirow{6}{*}{\tabincell{c}{PCA-RF\\\cite{machine5}}}& 0.6247 & 0.8850 & 0.6307 & 0.8867 & 0.6459 & 1 & 0.7426 & 0.8386 & 0.6684 & 0.8994 & 0.6494 & \multirow{6}{*}{\tabincell{c}{LSTM\\\cite{deep6}}}\\
		 & 0.6005 & 0.8719 & 0.6348 & 0.8893 & 0.6470 & 2 & 0.7775 & 0.7704 & 0.7763 & 0.9202 & 0.7729 &\\
		 & 0.6354 & 0.8715 & 0.6653	& 0.8948 & 0.6886 & 3 & 0.7185 & 0.7164	& 0.7144 & 0.9010 & 0.7099 &\\
		 & 0.6220 & 0.8577 & 0.6304	& 0.8829 & 0.6663 & 4 & 0.7480 & 0.7449 & 0.7153 & 0.9068 & 0.7232 &\\
		 & 0.6113 & 0.8722 & 0.5878	& 0.8730 & 0.6051 & 5 & 0.6622 & 0.5982 & 0.6031 & 0.8717 & 0.5791 &\\
		 \specialrule{0em}{0.5pt}{0.5pt}
		 \cline{2-12}
		 \specialrule{0em}{1pt}{1pt}
		 & 0.6188 & 0.8717 & 0.6298 & 0.8853 & 0.6506 & Average & 0.7298 & 0.7337 & 0.6955 & 0.8998 & 0.6869 &\\
\midrule
\multirow{6}{*}{\tabincell{c}{PCA-DT\\\cite{machine7}}}& 0.5845 & 0.5585 & 0.5732 & 0.8544 & 0.5648 & 1 & 0.7346 & 0.7271 & 0.7301 & 0.9031 & 0.7260 & \multirow{6}{*}{\tabincell{c}{CNN-RP\\\cite{raman_image2}}}\\
		 & 0.6702 & 0.6467 & 0.6643 & 0.8841 & 0.6532 & 2 & 0.6783 & 0.6635 & 0.6476 & 0.8826 & 0.6479 &\\
		 & 0.6488 & 0.6298 & 0.6364	& 0.8782 & 0.6302 & 3 & 0.6836 & 0.6863 & 0.6617 & 0.8853 & 0.6681 &\\
		 & 0.5952 & 0.5972 & 0.5643	& 0.8526 & 0.5772 & 4 & 0.7534 & 0.7417 & 0.7278 & 0.9094 & 0.7335 &\\
		 & 0.6032 & 0.6039 & 0.5960	& 0.8579 & 0.5981 & 5 & 0.6890 & 0.7027 & 0.6688 & 0.8887 & 0.6818 &\\
		 \specialrule{0em}{0.5pt}{0.5pt}
		 \cline{2-12}
		 \specialrule{0em}{1pt}{1pt}
		 & 0.6204 & 0.6072 & 0.6068 & 0.8654 & 0.6047 & Average & 0.7078 & 0.7042 & 0.6872 & 0.8938 & 0.6914 &\\
\midrule
\multirow{6}{*}{\tabincell{c}{PCA-KNN\\\cite{machine10}}}& 0.8043 & 0.8335 & 0.8105 & 0.9343 & 0.8187 & 1 & 0.8552 & 0.8717 & 0.8370 & 0.9461 & 0.8539 & \multirow{6}{*}{\tabincell{c}{CNN-GAF\\\cite{raman_image2}}}\\
		 & 0.8150 & 0.8421 & 0.8160 & 0.9343 & 0.8268 & 2 & 0.9115 & 0.9115 & 0.9017 & 0.9687 & 0.9065 &\\
		 & 0.7882 & 0.8211 & 0.7927	& 0.9288 & 0.8051 & 3 & 0.8713 & 0.8657 & 0.8544 & 0.9540 & 0.8600 &\\
		 & 0.8043 & 0.8344 & 0.8125	& 0.9301 & 0.8201 & 4 & 0.8660 & 0.8788 & 0.8623 & 0.9514 & 0.8704 &\\
		 & 0.7802 & 0.7947 & 0.7741	& 0.9234 & 0.7808 & 5 & 0.8767 & 0.8954 & 0.8733 & 0.9546 & 0.8842 &\\
		 \specialrule{0em}{0.5pt}{0.5pt}
		 \cline{2-12}
		 \specialrule{0em}{1pt}{1pt}
		 & 0.7984 & 0.8252 & 0.8012 & 0.9302 & 0.8103 & Average & 0.8761 & 0.8846 & 0.8657 & 0.9549 & 0.8750 &\\
\midrule
\multirow{6}{*}{\tabincell{c}{PCA-Adaboost\\\cite{machine7}}}& 0.5952 & 0.6041 & 0.6069 & 0.8570 & 0.6011 & 1 & 0.8311 & 0.8144 & 0.8244 & 0.9395 & 0.8184 & \multirow{6}{*}{\tabincell{c}{CNN-MTF\\\cite{raman_image2}}}\\
		 & 0.6676 & 0.7002 & 0.6509 & 0.8782 & 0.6689 & 2 & 0.8606 & 0.8573 & 0.8506 & 0.9494 & 0.8534 &\\
		 & 0.6649 & 0.6692 & 0.6384 & 0.8760 & 0.6466 & 3 & 0.8150 & 0.8358 & 0.7880 & 0.9317 & 0.8051 &\\
		 & 0.6139 & 0.6035 & 0.5538	& 0.8559 & 0.5661 & 4 & 0.8606 & 0.8440 & 0.8584 & 0.9497 & 0.8505 &\\
		 & 0.6166 & 0.6276 & 0.6373	& 0.8674 & 0.6240 & 5 & 0.8660 & 0.8748 & 0.8594 & 0.9491 & 0.8651 &\\
		 \specialrule{0em}{0.5pt}{0.5pt}
		 \cline{2-12}
		 \specialrule{0em}{1pt}{1pt}
		 & 0.6316 & 0.6409 & 0.6174 & 0.8669 & 0.6213 & Average & 0.8466 & 0.8452 & 0.8362 & 0.9438 & 0.8385 &\\
\midrule
\multirow{6}{*}{\tabincell{c}{PCA-SVM\\\cite{machine2}}}& 0.6836 & 0.8880 & 0.6859 & 0.9072 & 0.6878 & 1 & 0.9276 & 0.9358 & 0.9297 & 0.9731 & 0.9327 & \multirow{6}{*}{\tabincell{c}{M3S\\(Ours)}}\\
		 & 0.6810 & 0.8822 & 0.6822	& 0.9040 & 0.6701 & 2 & 0.9383 & 0.9400 & 0.9402 & 0.9777 & 0.9401 &\\
		 & 0.6783 & 0.8648 & 0.6912 & 0.9060 & 0.6887 & 3 & 0.9330 & 0.9298 & 0.9315 & 0.9755 & 0.9306 &\\
		 & 0.6702 & 0.8807 & 0.6566	& 0.8988 & 0.6576 & 4 & 0.9330 & 0.9505 & 0.9163 & 0.9747 & 0.9330 &\\
		 & 0.7105 & 0.8864 & 0.6737 & 0.9082 & 0.6777 & 5 & 0.9330 & 0.9333 & 0.9280 & 0.9748 & 0.9306 &\\
		 \specialrule{0em}{0.5pt}{0.5pt}
		 \cline{2-12}
		 \specialrule{0em}{1pt}{1pt}
		 & 0.6847 & 0.8804 & 0.6779 & 0.9048 & 0.6764 & Average & \textbf{0.9330} & \textbf{0.9379} & \textbf{0.9291} & \textbf{0.9752} & \textbf{0.9334} &\\
\bottomrule
\end{tabular}
}
\label{table2}
\end{table}

\subsubsection{Experimental Setup}
The input resolution of Raman images is 32 $\times$ 32 and 64 $\times$ 64, respectively. We use a two-branch CNN structure as the feature extractor with a learning rate of 0.001 in the SGD optimizer. 
The initial value of the weight matrix is 1, and the experiment is performed over 500 epochs until the loss value is converged.
All experiments are implemented using PyTorch 1.10.0, Python 3.8, and Cuda 11.3 framework on a single NVIDIA RTX 3090-24GB GPU, 10-core Intel(R) Xeon(R) Gold 5218R CPU @ 2.10GHz, and 64GB RAM.

\begin{figure}[t]
\centering\includegraphics[width=1\textwidth]{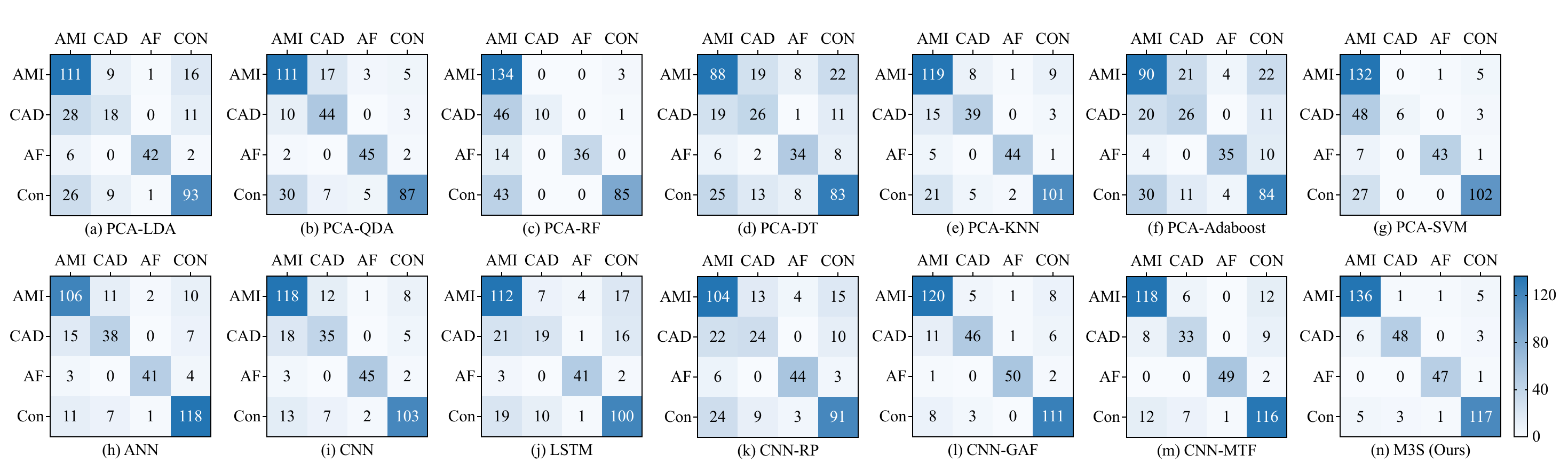}
\caption{The average confusion matrixs of our model and baselines with five random tests.} 
\label{SOTA-heatmap}
\end{figure}

\subsection{Diagnostic Performance (Q1)}
In order to verify the effectiveness of our model, we compare the performance of M3S with state-of-the-art algorithms. 
As shown in Table \ref{table2}, we use five random tests for all experiments and compare the results by averaging them. 
From the result, we can see that the fluctuations of the five test results of most algorithms are relatively small. At the same time, there is no obvious over-fitting or under-fitting phenomenon in the test results, which shows that the data collected in this paper can generate valuable results. 
Moreover, the M3S consistently outperforms most of the listed models.
It is reasonable as our model can discriminate the nuance features and generate a specific representation for subtypes, effectively improving its classification performance on CVD datasets. 
Specifically, the first 32 principal components can represent the original dataset better than other dimensions after PCA analysis, so these components are the input variables for the conventional machine learning methods. 
Based on these components, the conventional methods are able to classify the main subtypes of the CVD, and the PCA-KNN achieves the best classification accuracy. 
However, the RS data are usually infected by fluorescence and background noise, making it difficult for noise-sensitive machine learning algorithms to distinguish these similar spectroscopies accurately. 
And these methods cannot meet the needs of practical applications.  
On the contrary, the deep learning method can learn the features that are more robust to drift than the main components and achieve high classification accuracy. 
The results of these models have apparent improvement compared to conventional methods, demonstrating that encoding a single RS into a 2D spectral GAF image is more suitable for deep learning input. 
Although these methods reduce the impact of noise sensitivity, they lack consideration of the learning capacity from multi-scale and multi-modality data.

In view of this, we propose a multi-modality multi-scale method that can capture the extensive details from different forms and perspectives of data. 
The result in Table \ref{table2} significantly shows that our model has higher accuracy than conventional machine learning and deep learning methods, and it achieves the best results in classifying subtypes. 
The detailed performance for classifying subtypes of CVD is displayed in the confusion matrixs, as shown in Fig. \ref{SOTA-heatmap}. 
The entry in column $a$, row $b$ represents the quantity that test data from the output of the CNN are of class $a$, given a ground truth of class $b$.
These results illustrate that our model has higher accuracy and less possibility of misdiagnosis for distinguishing subtypes and control group data by combining the multi-scale and multi-modality modules. 

\subsection{Effectiveness of Multi-Scale Module (Q2)}

\begin{figure}[t]
\centering\includegraphics[width=0.95\textwidth]{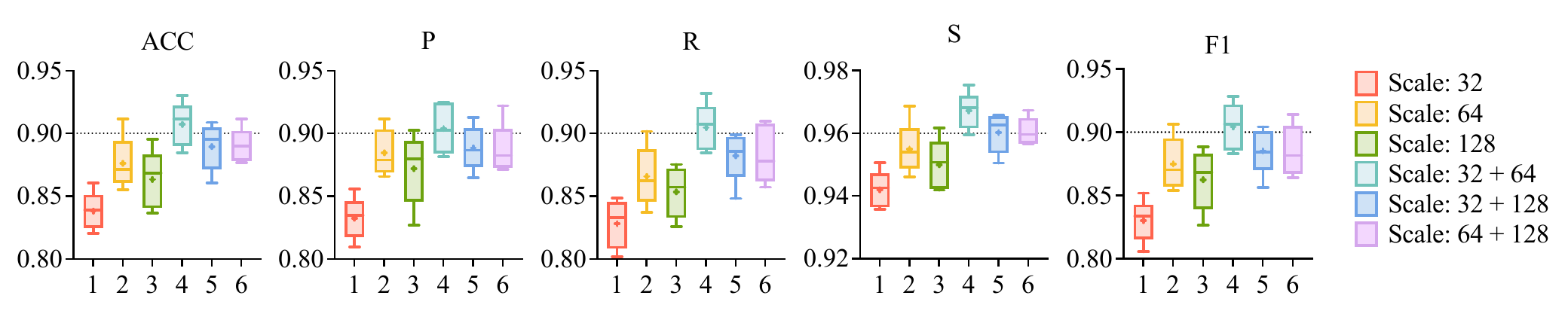}
\caption{Box and whisker plots between different scales with five random tests.} 
\label{multi-scale-image}
\end{figure}

To verify the effectiveness of the multi-scale module, we create five random test sets and record the metric results. 
We select three scales with good performance and combine them to form the multi-scale module, so there have six structures for generating GAF images. 
M3S uses the same feature extraction network and does not incorporate medical history information at this stage. 
As shown in Fig. \ref{multi-scale-image}, results of different scales are enclosed in each box, and the data distribution of different scales is relatively concentrated. The length of the box in the figure shows that the model is relatively stable in multiple test results, which reveals the robustness of the multi-scale module.  
Among these five metric evaluations, we find that the box of scale "32 + 64" is always in the highest position, indicating the effectiveness and stability of the multi-scale feature extraction module. 

\begin{table}[t]
\centering
\caption{Performance comparison of M3S between different scales without medical history information. The results are averaged with five random tests.}
\footnotesize
\setlength{\tabcolsep}{2mm}{
\begin{tabular}{c|ccccc}
\toprule
Scale & ACC & P & R & S & F1 \\
\midrule
\midrule
        32 & 0.8381	& 0.8324 & 0.8280 & 0.9420 & 0.8301 \\
        64 & 0.8761	& 0.8846 & 0.8657 & 0.9549 & 0.8750 \\
        128 & 0.8633 & 0.8720 & 0.8534 & 0.9500	& 0.8625 \\
        32 + 64 & \textbf{0.9072} & \textbf{0.9039} & \textbf{0.9047} & \textbf{0.9671} & \textbf{0.9043} \\
        32 + 128 & 0.8895 & 0.8884 & 0.8822 & 0.9602 & 0.8853 \\
        64 + 128 & 0.8901 & 0.8871 & 0.8835 & 0.9606 & 0.8852 \\
\bottomrule
\end{tabular}
}
\label{table3}
\end{table}

Table \ref{table3} shows the average results of this module, and the methods combining multi-scale images are better than the single-scale model, especially the combination of "32 + 64". 
It outperforms other single-scale and multi-scale results, illustrating that using a combination of 32 and 64 scales is the best choice to extract recognizable features. 
As we have mentioned in Q1, the conventional methods can get the best performance by using 32 principal components. 
And the results from deep learning methods also show that it is sufficient for subtypes classification using 32 and 64 feature points. 
It demonstrates that the multi-scale structure can automatically discover and screen these crucial features to provide an excellent preliminary prediction at this stage. 

\subsection{Benefits of Multi-Modality Module (Q3)}
\begin{figure}[t]
\centering\includegraphics[width=1\textwidth]{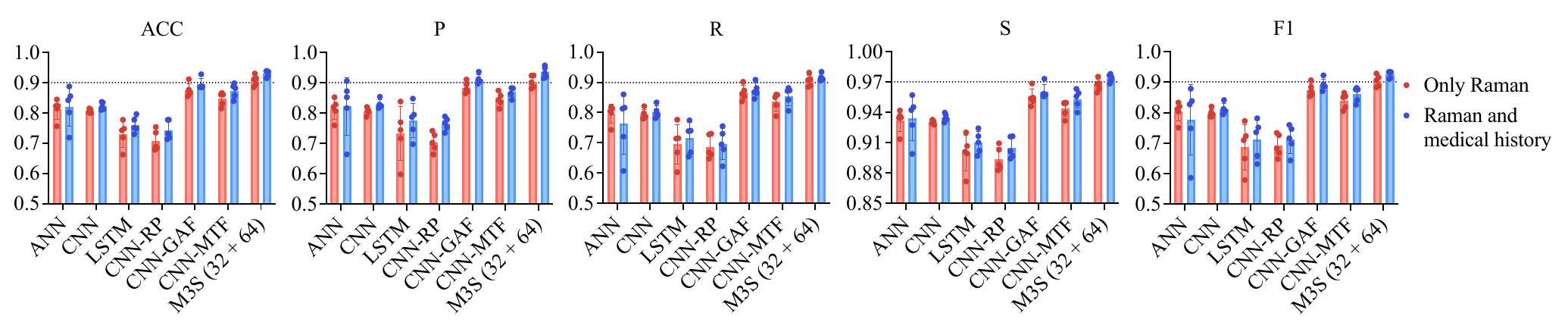}
\caption{The five random tests of deep learning methods and M3S fusing medical history information by the fixed weight matrix.} 
\label{multi-modality-image}
\end{figure}

In this subsection, we intend to investigate whether considering the multi-modality data is an effective way to help the model improve classification performance than only using Raman. 
We choose deep learning models similar to our model for fairness comparison. And we use the fixed weight matrix (the weight ratio of RS and medical history is 9:1) to ensure that the structures of other models are not modified. 
Fig. \ref{multi-modality-image} shows the changes in evaluation metrics before and after combining various deep learning models with medical history information. 
Most of the models show significant improvement in classification performance with the help of multi-modality, especially for the evaluation of ACC. 
The results of ANN and LSTM fluctuate considerably. 
In contrast, the results of our model can stably fluctuate with a small range in blue bars, which shows the robustness of the multi-modality module. 

\begin{table}[t]
\centering
\caption{Performance comparison of deep learning methods and M3S when combined with medical history information by the fixed weight matrix. The results are averaged with five random tests.}
\footnotesize
\setlength{\tabcolsep}{0.48mm}{
\begin{tabular}{c|ccccc|cc}
\toprule
Method & ACC & P & R & S & F1 & FLOPs & Params \\
\midrule
\midrule
        \tabincell{c}{ANN} & 0.8209 & 0.8226 & 0.7634 & 0.9342 & 0.7758 & \textbf{0.70M} & 0.70M \\
        \tabincell{c}{CNN} & 0.8225 & 0.8293 & 0.8021 & 0.9346 & 0.8116 & 38.26M & 0.13M \\
        \tabincell{c}{LSTM} & 0.7603 & 0.7750 & 0.7154 & 0.9100 & 0.7117 & 3.16M & 3.15M\\
        \tabincell{c}{CNN-RP} & 0.7437 & 0.7650 & 0.6964 & 0.9050 & 0.7128 & 2.89M & \textbf{0.04M} \\
        \tabincell{c}{CNN-GAF} & 0.8944 & 0.9118 & 0.8732 & 0.9605 & 0.8921 & 2.89M & \textbf{0.04M} \\
        \tabincell{c}{CNN-MTF} & 0.8729 & 0.8711 & 0.8548 & 0.9533 & 0.8603 & 2.89M & \textbf{0.04M} \\
\midrule
        \tabincell{c}{M3S \\(32 + 64)} & \textbf{0.9298} & \textbf{0.9353} & \textbf{0.9152} & \textbf{0.9740} & \textbf{0.9251} & 6.75M & 0.12M \\
\bottomrule
\end{tabular}
}
\label{table4}
\end{table}

Second, Table \ref{table4} shows the average results of different deep learning methods. 
It shows that their classification performance has improved compared with Table \ref{table2} when combining the medical history information with Raman data. 
And compared with other methods, our model combining multi-modality data by the fixed weight matrix has a better classification ability in each indicator. 
In addition, we separately calculated the number of floating point operations and the number of parameters for these deep learning algorithms for comparing the model complexity. 
Compared with the algorithms with the lowest FLOPs and Params, the computational complexity of our model slightly increases, but the classification performance is significantly improved. 
According to these results, it can be concluded that our model does better in pursuit of a trade-off between accuracy and computational efficiency, which is a crucial goal in medical diagnosis tasks. 
In summary, it powerfully illustrates the importance and necessity of combining multi-modality to classify subtypes of CVD. 

\subsection{In Deep Analysis of M3S (Q4)}
For an in-depth analysis of M3S, we enumerate the classification results from different components of M3S and visualize the weight matrix, which can help us analyze the advantages and significances of our models from both quantitative and qualitative perspectives. 

\begin{table}[t]
\centering
\caption{Performance comparison of M3S between different scales and fusion methods. Group (a) is obtained by the fixed weight matrix, and group (b) is obtained by the adaptive weight matrix. The results are averaged with five random tests.}
\footnotesize
\setlength{\tabcolsep}{1.4mm}{
\begin{tabular}{cc|ccccc}
\toprule
Group & Scale & ACC & P & R & S & F1 \\
\midrule
\midrule
        \multirow{6}{*}{(a)} & 32 & 0.8660 & 0.8617 & 0.8509 & 0.9515 & 0.8562 \\
        & 64 & 0.8944 & 0.9118 & 0.8732	& 0.9605 & 0.8921 \\
        & 128 & 0.8858 & 0.9110 & 0.8589 & 0.9568 & 0.8841 \\
        & 32 + 64 & 0.9298 & 0.9353 & 0.9152 & 0.9740 & 0.9251 \\
        & 32 + 128 & 0.9110 & 0.9236 & 0.8914 & 0.9665 & 0.9072 \\
        & 64 + 128 & 0.9158 & 0.9237 & 0.8950 & 0.9685 & 0.9091 \\
\midrule
        \multirow{6}{*}{(b)} & 32 & 0.8874 & 0.8845 & 0.8720 & 0.9599 & 0.8781 \\
        & 64 & 0.9233 & 0.9290 & 0.9143 & 0.9720 & 0.9216 \\
        & 128 & 0.9131 & 0.9140 & 0.9084 & 0.9687 & 0.9111 \\
        & 32 + 64 & \textbf{0.9330} & \textbf{0.9379} & \textbf{0.9291} & \textbf{0.9752} & \textbf{0.9334} \\
        & 32 + 128 & 0.9191 & 0.9145 & 0.9136 & 0.9714 & 0.9140 \\
        & 64 + 128 & 0.9056 & 0.9062 & 0.8954 & 0.9656 & 0.9007 \\
\bottomrule
\end{tabular}
}
\label{table5}
\end{table}

Firstly, will the adaptive weights learned by M3S perform better than expert experiments? 
It has been mentioned above that the fixed weight ratio of RS and medical history information is 9:1, which is an empirical value given by medical experts. 
But experience is not always the best teacher, and we can not confirm it is the best-fixed ratio for fusion. 
At the same time, it is hard for us to discover the relationship between medical histories, RS data, and subtypes, which brings an ambiguous explanation for the diagnosis.
Therefore,  we design a learnable adaptive weight matrix that can assign mixing ratios of multi-modality between different medical histories and subtypes. 
Table \ref{table5} shows the classification performance of M3S using fixed and adaptive weights to fuse RS and medical history information. 
The adaptive weights can perform better than fixed weights by comparing multiple metrics. 
Moreover, Table \ref{table6} shows the ablation studies of our model, and we compare the single-scale, multi-scale, fixed-weight, and adaptive weight approaches, respectively. 
The classification accuracy of M3S is 93.3\%, which is 5.69\% higher than the method using Raman image with a single scale. 
It demonstrates the advantages of the M3S by combining multi-scale and multi-modality modules from a quantitative perspective. 

\begin{table}[t]
\centering
\caption{Evaluation of different components effectiveness with M3S for subtypes classification of CVD. Accuracy (\%).}
\footnotesize
\setlength{\tabcolsep}{2mm}{
\begin{tabular}{cccc|cc}
\toprule
\tabincell{c}{scale\\64} & \tabincell{c}{scale\\32 + 64} & \tabincell{c}{fixed\\weight} & \tabincell{c}{adaptive\\weight} & ACC & Gain ($\uparrow$) \\
\midrule
\midrule
        \ding{52} &  &  &  & 87.61\% & - \\
         & \ding{52} &  &  & 90.72\% & 3.11\% \\
         & \ding{52} & \ding{52} &  & 92.98\% & 5.37\% \\
         & \ding{52} &  & \ding{52} & \textbf{93.30\%} & \textbf{5.69\%}\\
\bottomrule
\end{tabular}
}
\label{table6}
\end{table}

\begin{figure}[t]
\centering\includegraphics[width=0.7\textwidth]{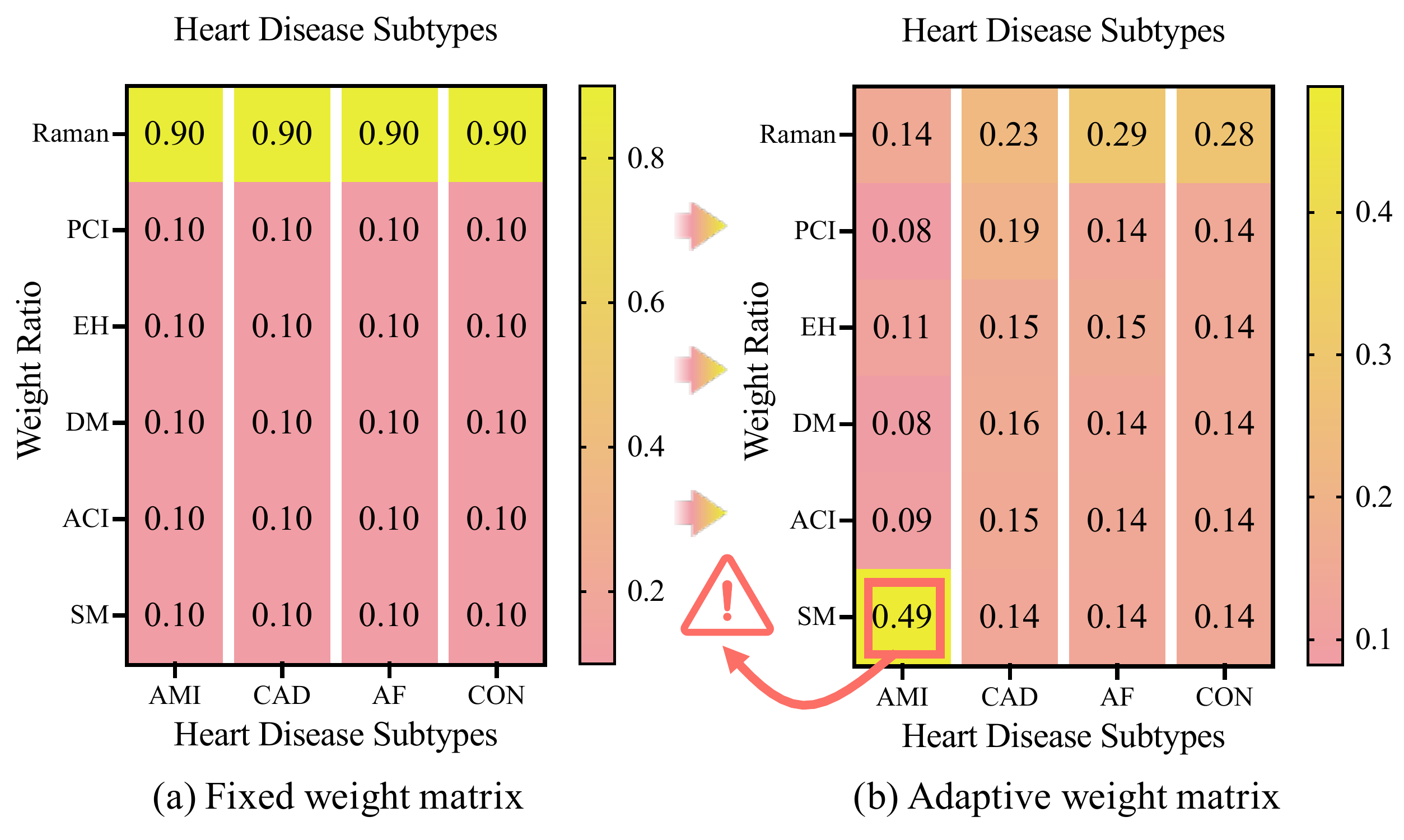}
\caption{Interpretation comparison of M3S between fixed weight matrix and adaptive weight matrix.} 
\label{mixing-ratio-image}
\end{figure}

Secondly, what is the potential relationship between subtypes and medical histories?  
We visualized the adaptive weight matrix parameters and activated them using the softmax function. 
As shown in Fig. \ref{mixing-ratio-image}, the adaptive weight matrix has different mixing ratios for subtypes and medical histories compared to the fixed weight matrix. 
To be specific, the influence of smoking on the diagnosis of AMI is more significant than results from RS, which is also in line with the actual clinical situation. 
Based on the value of this weight matrix, the model can clearly explain to the doctor why the patient is diagnosed with AMI rather than CAD or AF. 
Therefore, the adaptive weight matrix can reflect the profound relationship between medical histories and subtypes and allows us to understand the influencing factors that have not yet been discovered to make a more scientific diagnosis. 
Compared with the empirical value from experts, the mixing ratios learned by the model have better performance and interpretability. 
In summary, using the weight matrix to combine RS and medical history information has great significance for the subtypes classification of CVD.

\begin{figure}[t]
\centering\includegraphics[width=0.7\textwidth]{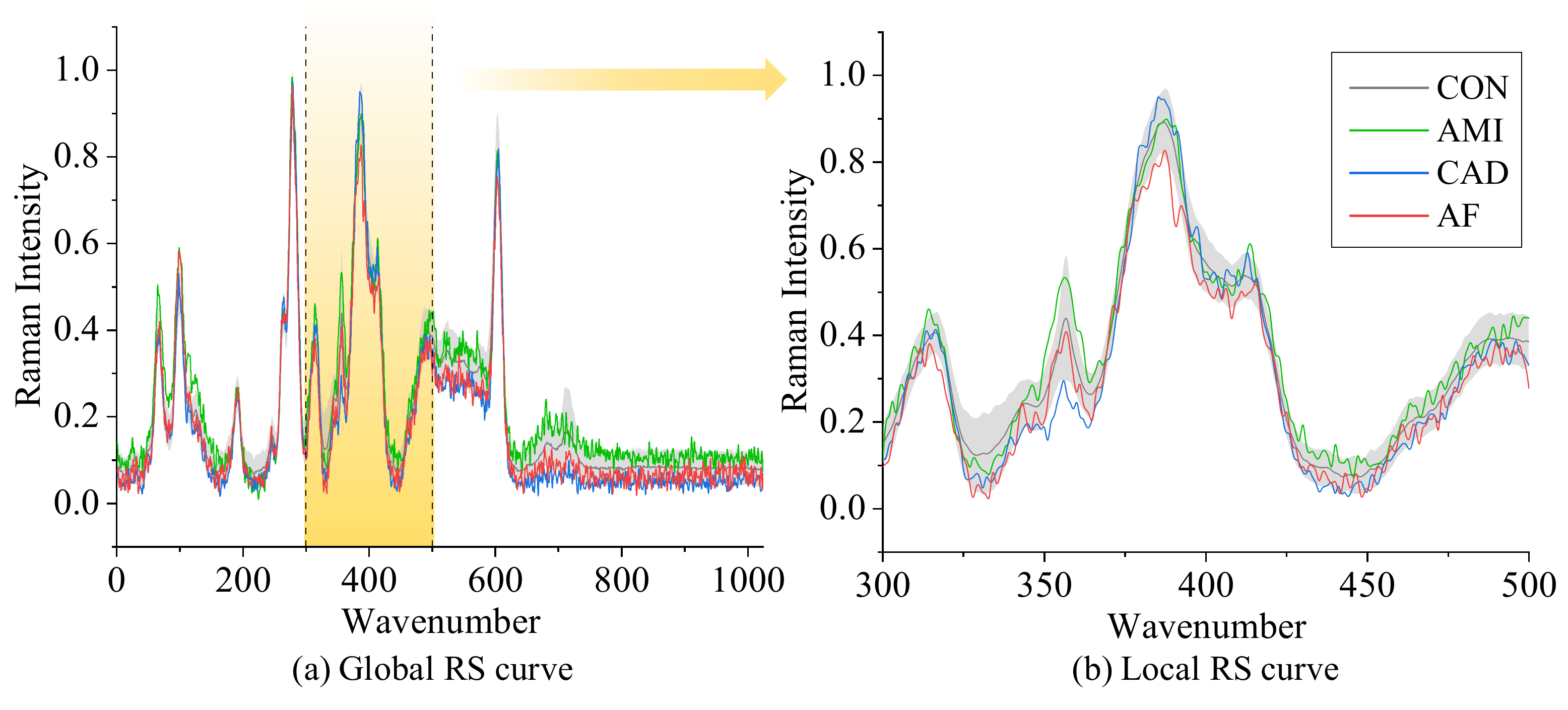}
\caption{Diagram of misclassified samples and all CON samples in the test set. The cases of AMI, CAD, and AF are predicted as CON. The gray-shaded part is obtained by calculating the mean and standard deviation of all CON cases in the test set.} 
\label{error-analysis}
\end{figure}

Finally, as with the majority of studies, the design of the current study is subject to limitations.
One of the most crucial issues is the indistinguishability of remarkably alike spectroscopy of different subtypes.
According to Fig. \ref{SOTA-heatmap} (n), 26 sequences are mispredicted by M3S, while 18 sequences are related to the CON. 
Therefore, we visualize all CON data in the test set and display them as gray lines and shading in terms of mean and standard deviation. 
The RS distribution curves of the three failed subtype data with the test set of CON are shown in Fig. \ref{error-analysis} (a).
To observe more clearly, we zoom these RS data on the region from 300 to 500, as shown in Fig. \ref{error-analysis} (b).
There is an obvious discovery that the spectrums of the subtype diseases are mainly distributed on the edge of the CON set, but many points still overlap with CON. 
One of the reasons for this is that since a sample will be collected multiple times for testing, the sample with less lesion information will be very close to the CON.
In addition, the complexity optimization of the model and the expansion of the data set scale are the directions of our future research. 

\section{Related Work}
This section briefly reviews previous work on RS-based intelligent disease analysis and diagnosis methods, including conventional machine learning and deep learning methods.

\subsection{Conventional Machine Learning Methods}
Typically, these conventional methods use PCA to perform dimension reduction from the RS sequence to construct classifiers for distinguishing patients from normal controls.
For example, \cite{machine8} and \cite{machine9} use PCA to get principal components from RS sequences of CHD, and then LDA is performed in the PCA subspace for distinction. 
\cite{machine3} also give a new idea by using this method for the detection of chronic kidney disease.
\cite{machine6} propose a PCA-QDA method to distinguish grade I and grade II of brain tumors by RS in a 3D space.
Further, \cite{machine5} combinate PCA and RF to analyze glioma biopsies on fresh tissue samples.
\cite{machine7} make some effort to classify papillary thyroid carcinoma and papillary microcarcinoma by using DT, RF, and Adaboost on RS.
\cite{machine10} propose to build a detection model for high renin hypertension by using PCA and KNN.
Finally, SVM is the most popular and successful algorithm for classifying RS sequences, which has been widely used to diagnose diabetes mellitus \cite{machine1}, breast cancers \cite{machine2}, and glioma biopsies \cite{machine4}.
However, these methods are sensitive to noise in the RS data, which makes them hard to use in real-world applications.

\subsection{Deep Learning Methods}
Recently, deep learning methods have been explored to extract high-level RS features in a data-driven manner for disease diagnosis. 
For example, \cite{deep4} apply a deep backpropagation (BP) neural network on fluorescence imaging and RS to accurately diagnose breast cancer.
With the development of CNN, CNN-based methods for disease diagnosis on RS have made significant progress \cite{deep8}.
\cite{deep1} use CNN for rapid identification of pathogenic bacteria.
\cite{deep2} perform CNN for prostate cancer detection on RS of extracellular vesicles.
\cite{deep3} propose a precise cancer detection method that combines RS and CNN with independent inputs.
It is worth mentioning that some researchers propose to transform RS data into Raman figures by the GAF, RP, and MTF, then combine them with CNN to improve the accuracy \cite{raman_image1,raman_image2}.
In addition, the RNN model has also been applied to RS data for disease diagnosis. \cite{deep6} and \cite{deep7} build innovative models for analyzing marine pathogens and hepatitis B by LTSM, which is a form of RNN.
Although the generalizability of these methods would probably lead to more accurate classification, they still lack the ability to classify disease subtypes, e.g., AMI, CAD, and AF.

The core idea of M3S is inspired to solve \textbf{\emph{challenge I}} and \textbf{\emph{challenge II}} for classifying subtypes of CVD on RS data. 
Thus, we first design the multi-scale feature extraction module that can convert sequences to various resolutions of images and extract their features by a two-branch CNN structure.
Then, the multi-modality data fusion module is designed to combine data between RS and medical history information by two matrixes. 

\section{Conclusion}
This paper proposes a novel M3S model with two modules, which implements the diagnosis process of CVD subtypes with RS data and medical histories. 
GAF is used to convert RS to images with various resolutions, and then we design a multi-scale CNN structure for feature extraction and representation. 
The probability matrix expresses the interaction between medical histories and subtypes, and the weight matrix is used to assign the mixing ratio for multi-modality data.
The study provides sound experimental evidence that M3S has ultimate performance in the classification of CVD subtypes, and it outperforms the state-of-the-art methods. 
In general, it demonstrates that combining the 2D Raman images and medical histories is a possible way to diagnose CVD accurately. 
Most importantly, the model proposed in this paper has practical novelties. This method can potentially replace some invasive methods in detecting CVD subtypes in the clinic with a non-invasive approach. At the same time, it can give fast and accurate diagnostic results and assist doctors in implementing scientific surgical plans.
In the follow-up work, an optimized M3S structure will be adopted and applied to other disease subtypes for classification to verify its applicability further.

\section*{Acknowledgments}
This work is partially supported in part by the International Cooperation Project under grant No. 20220402009GH; the National Natural Science Foundation of China under grants Nos. 61976102, and U19A2065; the National Key R\&D Program of China under grants Nos. 2021ZD0112501 and 2021ZD0112502; the project of Shanghai Tenth People's Hospital (No. YNCR2A013); the China Scholarship Council Project (202206170090).

\bibliographystyle{unsrt}  
\bibliography{references}

\end{document}